\documentclass[conference]{IEEEtran}
\IEEEoverridecommandlockouts
\usepackage{cite}
\usepackage{amsmath,amssymb,amsfonts}
\usepackage{graphicx}
\usepackage{textcomp}
\usepackage{xcolor}
\def\BibTeX{{\rm B\kern-.05em{\sc i\kern-.025em b}\kern-.08em
    T\kern-.1667em\lower.7ex\hbox{E}\kern-.125emX}}

\usepackage{amsthm,mathtools}
\usepackage{algorithm}
\usepackage{algorithmic}
\usepackage{booktabs}
\usepackage{array}
\usepackage{url}
\usepackage{caption}
\usepackage{multirow}

\newtheorem{remark}{Remark}

\newcommand{\tr}{\operatorname{tr}}

\newcommand{\Hs}{\mathcal{H}}

\newcommand{\C}{\mathbb{C}}
\newcommand{\E}{\mathbb{E}}

\begin{document}

\title{Online Estimation of Partial Transpose Moments\\via Fast Classical Updates
}

\author{\IEEEauthorblockN{Hyunho Cha and Jungwoo Lee}
\IEEEauthorblockA{\textit{NextQuantum and Department of Electrical and Computer Engineering} \\
\textit{Seoul National University}\\
Seoul, Republic of Korea \\
\texttt{\{ovalavo, junglee\}@snu.ac.kr}}
}

\maketitle


\begin{abstract}
Partial-transpose (PT) moments are among the most practically relevant nonlinear quantities accessible from local Pauli classical shadows, because they directly underpin mixed-state entanglement certification and recent PT-moment-based phase diagnostics.
The online framework of Marso \emph{et al.} rewrote the exact PT-moment statistic into a fixed-memory recurrence that updates a small collection of accumulated matrices after each new shadow snapshot.
Its update cost is independent of the shot number, but each step treats the incoming partially transposed snapshot as a generic dense matrix. Therefore, the arithmetic cost scales cubically with the dimension of the Hilbert space.
We show that the same estimator can be updated exactly in subcubic time per shot while retaining the same memory. The key point is that the accumulated matrices become dense, but the fresh partially transposed snapshot still factorizes into local factors. Right-multiplication by that factorized snapshot can therefore be executed by exact column-pair sweeps.
For the second PT moment, we further optimize the process by utilizing a Pauli basis update.
\end{abstract}


\section{Introduction}

Shadow tomography asks for a compact classical record from which many properties of a quantum state can be predicted. Aaronson formulated the general shadow-tomography problem in information-theoretic terms \cite{aaronson2018shadow}, and Huang \emph{et al.} introduced the experimentally practical classical-shadow framework based on randomized measurements and a simple inversion map \cite{huang2020predicting}. Randomized measurements and classical shadows have since become a standard tool for near-term platforms \cite{elben2023randomized,cieslinski2024analysing}.

A particularly important application is mixed-state entanglement detection. Elben \emph{et al.} showed that moments of a partially transposed density matrix can be estimated from local randomized measurements \cite{elben2020mixed}. Neven \emph{et al.} then developed a hierarchy of PT-moment inequalities whose union reproduces the Peres--Horodecki positive-partial-transpose criterion \cite{neven2021symmetry}. PT moments have also been used to organize entanglement phase diagrams \cite{carrasco2024entanglement}, and recent work continues to expand the family of PT-moment-based witnesses \cite{miller2026detecting}. More general trace-polynomial witnesses accessible from randomized measurements were developed in \cite{rico2024entanglement}, while \cite{rico2025state} showed how to tailor such witnesses without sacrificing randomized-measurement implementability.

This work concerns a classical bottleneck in the online processing loop. In hybrid experiments, shots arrive sequentially, and classical post-processing can often be overlapped with reset and control latency. Motivated by this architecture, Marso \emph{et al.} recently introduced online PT-moment estimators from local Pauli shadows \cite{marso2026online}. Their reconstruction-free estimator is memory-light but has a combinatorial dependence on the number of previous shots. On the other hand, their reconstruction-based estimator stores a fixed set of accumulated matrices, discards the fresh shot immediately after processing, and updates in time independent of the number of past shots. This fixed-memory, shot-independent regime is especially attractive when one wants high PT moments but cannot retain the full measurement history.

Its remaining drawback is arithmetic cost. If $d=2^n$, the recurrence updates $m$ dense $d\times d$ matrices by multiplying each of them with the newly arrived partially transposed shadow snapshot, which yields a standard-cost bound of $O(md^3)$ per shot. Our contribution is to accelerate this reconstruction-based estimator without changing the estimator itself.
The main observation is that, even though the stored matrices become dense, the newest right factor can still be applied in $n$ local sweeps, and that one-sided structure reduces the cost of every update.
We also give a separate specialization for the second PT moment, where the estimator can be maintained through the Pauli coefficients, reducing the per-shot arithmetic further to $O(d)$.

\section{Preliminaries}

\subsection{Basic concepts}

A single qubit is described by the two-dimensional complex Hilbert space $\C^2$. An $n$-qubit system lives in the tensor-product space $\Hs=(\C^2)^{\otimes n}$ with dimension $d=2^n$.
The computational basis of $\Hs$ is indexed by bit strings $x=(x_1,\ldots,x_n)\in\{0,1\}^n$. Unless stated otherwise, every transpose $M^T$ in this paper is taken in this computational basis.

A quantum state on $\Hs$ is represented by a density matrix $\rho$, that is, a Hermitian positive-semidefinite operator with unit trace:
$
\rho\succeq 0
$
and
$
\tr(\rho)=1.
$
Here $M\succeq 0$ means that all eigenvalues of $M$ are nonnegative. Pure states are the special case in which $\rho$ has rank one, while mixed states are convex combinations of pure states. If $O$ is a Hermitian operator, then its expectation value on $\rho$ is
$
\tr(\rho O).
$
More generally, a quantum measurement with outcomes indexed by $k$ is described by a positive operator-valued measure (POVM), that is, a collection of operators $\{M_k\}_k$ satisfying
$
M_k \succeq 0
$
for all $k$ and
$
\sum_k M_k = I.
$
When this measurement is applied to the state $\rho$, outcome $k$ occurs with probability
$
\tr(\rho M_k).
$
A projective measurement is the special case in which the POVM elements are orthogonal projectors, written $\{\Pi_k\}_k$.

The single-qubit operators used throughout this paper are the Pauli matrices $I = \left( \begin{smallmatrix} 1 & 0 \\ 0 & 1 \end{smallmatrix} \right)$, $X = \left( \begin{smallmatrix} 0 & 1 \\ 1 & 0 \end{smallmatrix} \right)$, $Y = \left( \begin{smallmatrix} 0 & -i \\ i & 0 \end{smallmatrix} \right)$, and $Z = \left( \begin{smallmatrix} 1 & 0 \\ 0 & -1 \end{smallmatrix} \right)$.
These matrices are Hermitian and have eigenvalues in $\{\pm 1\}$. On an $n$-qubit system, local operators are combined by tensor product. In particular, a Pauli string is an operator of the form
$
P_1\otimes\cdots\otimes P_n,
$
where
$
P_j\in\{I,X,Y,Z\}.
$

Fix a bipartition $[n]=A\sqcup B$. A state $\rho$ is \emph{separable} across $A|B$ if it can be written as
$
\rho=\sum_{\ell} p_\ell\, \rho_A^{(\ell)}\otimes \rho_B^{(\ell)}
$
with
$
p_\ell\ge 0
$
and
$
\sum_\ell p_\ell=1,
$
where each $\rho_A^{(\ell)}$ and $\rho_B^{(\ell)}$ is a density matrix on the corresponding subsystem. If no such decomposition exists, then $\rho$ is entangled across $A|B$.

\subsection{PT moments and entanglement certificates}

Let $\Hs=(\C^2)^{\otimes n}$ be the Hilbert space of an $n$-qubit state $\rho$, and let $d=2^n$. Fix a bipartition $[n]=A\sqcup B$. The partial transpose on subsystem $B$, denoted by $\Gamma_B$, is the linear map that transposes the tensor factors belonging to $B$ and leaves the tensor factors in $A$ unchanged. For a simple tensor,
\[
(M_1\otimes\cdots\otimes M_n)^{\Gamma_B}
=
\bigotimes_{j=1}^n M_j',
\quad
M_j'=
\begin{cases}
M_j,& j\notin B,\\
M_j^T,& j\in B.
\end{cases}
\]
We extend this definition by linearity to all operators on $\Hs$.

For an integer $m\ge 1$, the $m$-th PT moment is
\begin{equation}
p_m(\rho)=\tr[(\rho^{\Gamma_B})^m].
\label{eq:ptmoment}
\end{equation}
Because $\rho^{\Gamma_B}$ is Hermitian, its eigenvalues $\lambda_1,\ldots,\lambda_d$ are real and the PT moments are the corresponding power sums,
$
p_m(\rho)=\sum_{i=1}^d \lambda_i^m.
$
If $\rho$ is separable then $\rho^{\Gamma_B}\succeq 0$ by the Peres--Horodecki criterion \cite{peres1996separability,horodecki2001separability}. Therefore every elementary symmetric polynomial of the spectrum,
$
e_k(\lambda_1,\ldots,\lambda_d)
=
\sum_{1\le i_1<\cdots<i_k\le d}\lambda_{i_1}\cdots\lambda_{i_k},
$
must be nonnegative on separable states. To see the implication explicitly, note that separability implies $\rho^{\Gamma_B}\succeq 0$, hence every eigenvalue $\lambda_i$ is nonnegative. Every monomial appearing in $e_k(\lambda_1,\ldots,\lambda_d)$ is then nonnegative, so the whole sum is nonnegative as well. Therefore a negative value of any $e_k$ is incompatible with separability and certifies entanglement. In practice the $e_k$ are reconstructed from the PT moments by the Newton-Girard identities,
\begin{equation}
ke_k=\sum_{r=1}^k (-1)^{r-1}e_{k-r}p_r,
\qquad e_0=1,
\label{eq:newtongirard}
\end{equation}
which is the route used in \cite{neven2021symmetry,marso2026online}. Estimating PT moments online therefore means estimating an experimentally accessible entanglement witness online. For the purposes of this paper, the operational point is that the nonlinear witness construction sits entirely on top of the moment-estimation layer. Once the moments $p_1,\ldots,p_m$ have been updated, the conversion to the coefficients $e_k$ requires only low-dimensional scalar arithmetic through Eq.~\eqref{eq:newtongirard}. Consequently, in the fixed-memory online architecture of \cite{marso2026online}, the matrix recurrence is the dominant controller-side cost. Accelerating that recurrence therefore accelerates online entanglement certification.

\subsection{Local Pauli classical shadows}

We now specify the measurement model. On shot $t\in\{1,2,\ldots\}$ and qubit $j\in[n]$, choose
$
P_{t,j}\in\{X,Y,Z\}
$
uniformly at random and independently across both $t$ and $j$, measure in that basis, and record the outcome bit $b_{t,j}\in\{0,1\}$. It is convenient to collect these local data into the strings
$
P_t=(P_{t,1},\ldots,P_{t,n})\in\{X,Y,Z\}^n
$
and
$
b_t=(b_{t,1},\ldots,b_{t,n})\in\{0,1\}^n.
$
For $P\in\{X,Y,Z\}$ and $b\in\{0,1\}$, let
$
\Pi_{P,b}=\frac12(I+(-1)^bP)
$
be the projector onto the eigenstate of $P$ with eigenvalue $(-1)^b$. The $n$-qubit classical-shadow snapshot on shot $t$ is
\[
\hat\rho_t
=
\bigotimes_{j=1}^n (3\Pi_{P_{t,j},b_{t,j}}-I)
=
\bigotimes_{j=1}^n\left(\frac12 I+\frac32(-1)^{b_{t,j}}P_{t,j}\right).
\]

For one qubit, define
\[
\Phi_1(M)=\frac13\sum_{P\in\{X,Y,Z\}}\sum_{b=0}^1 \tr(M\Pi_{P,b})(3\Pi_{P,b}-I).
\]
Since $I,X,Y,Z$ form a basis of $2\times 2$ matrices, it suffices to evaluate $\Phi_1$ on that basis. For $M=I$,
$
\Phi_1(I)=\frac13\sum_{P\in\{X,Y,Z\}}\sum_{b=0}^1(3\Pi_{P,b}-I)
=\frac13\sum_{P\in\{X,Y,Z\}} I
= I,
$
because $\Pi_{P,0}+\Pi_{P,1}=I$ for every Pauli axis. For $M=Z$, only the $Z$ basis contributes, since $\tr(Z\Pi_{X,b})=\tr(Z\Pi_{Y,b})=0$ while $\tr(Z\Pi_{Z,b})=(-1)^b$. Hence
$
\Phi_1(Z)=\frac13\sum_{b=0}^1 (-1)^b(3\Pi_{Z,b}-I)
=\Pi_{Z,0}-\Pi_{Z,1}=Z.
$
The same calculation with the Pauli axes permuted gives $\Phi_1(X)=X$ and $\Phi_1(Y)=Y$. Thus $\Phi_1$ is the identity on all one-qubit operators. If $\Phi_n$ denotes the corresponding $n$-qubit averaging map, then the local product structure of the protocol implies $\Phi_n=\Phi_1^{\otimes n}$ on Pauli strings and hence on all operators. Tensorizing this identity over $n$ sites gives the usual local-Pauli reconstruction formula
$
\E[\hat\rho_t]=\rho.
$

The same local structure drives our new algorithm. For PT moments, the relevant object is not $\hat\rho_t$ itself but its partial transpose
$
S_t=(\hat\rho_t)^{\Gamma_B}.
$
The effect of $\Gamma_B$ on a single local factor follows from $X^T=X$, $Z^T=Z$, and $Y^T=-Y$.
Define the bit $\chi_{t,j} = 1$ if $j \in B$ and $P_{t,j} = Y$, and $\chi_{t,j} = 0$ otherwise.
If $j\notin B$, or if $j\in B$ but $P_{t,j}\in\{X,Z\}$, then the local factor is unchanged by transpose. If $j\in B$ and $P_{t,j}=Y$, then
$
\left(\frac12 I+\frac32(-1)^{b_{t,j}}Y\right)^T
=
\frac12 I-\frac32(-1)^{b_{t,j}}Y
=
\frac12 I+\frac32(-1)^{b_{t,j}\oplus 1}Y.
$
Thus partial transpose simply toggles the recorded bit on those $Y$-measured qubits that belong to $B$. As a result, $S_t$ retains the tensor-product form
\begin{equation}
S_t=\bigotimes_{j=1}^n G_{t,j},
\label{eq:factorized}
\end{equation}
where
$
G_{t,j}=
\frac12 I+\frac32(-1)^{b_{t,j}\oplus\chi_{t,j}}P_{t,j}.
$
Since partial transpose is linear,
$
\E[S_t]=\E[(\hat\rho_t)^{\Gamma_B}]=\rho^{\Gamma_B}.
$

\section{The online estimator and the bottleneck}

\subsection{The unbiased PT-moment estimator}

Because each $S_t$ is an unbiased estimator of $\rho^{\Gamma_B}$, the natural estimator for Eq.~\eqref{eq:ptmoment} is obtained by averaging trace products over all $m$-tuples of distinct shots:
\begin{equation}
\widehat p_m^{(N)}
=
\binom{N}{m}^{-1}
\sum_{1\le s_1<\cdots<s_m\le N}
\tr(S_{s_1}S_{s_2}\cdots S_{s_m}).
\label{eq:ustat}
\end{equation}
This is the $U$-statistic used in \cite{marso2026online}.

Fix one increasing $m$-tuple $s_1<\cdots<s_m$. The snapshots $S_{s_1},\ldots,S_{s_m}$ are independent because they arise from different shots. Therefore,
$
\E[S_{s_1}S_{s_2}\cdots S_{s_m}]
=
\E[S_{s_1}]\,\E[S_{s_2}]\cdots \E[S_{s_m}]
=
(\rho^{\Gamma_B})^m.
$
Taking traces and using linearity gives
$
\E[\tr(S_{s_1}\cdots S_{s_m})]
=
\tr[(\rho^{\Gamma_B})^m]
=
p_m(\rho).
$
Every summand in Eq.~\eqref{eq:ustat} has the same expectation, and there are exactly $\binom{N}{m}$ summands, so
$
\E[\widehat p_m^{(N)}]=p_m(\rho).
$
The estimator in Eq.~\eqref{eq:ustat} is unbiased, but in its raw form it is not yet practical as an online algorithm because the number of tuples grows combinatorially with $N$.

Specifically, the online framework of \cite{marso2026online} contains two genuinely different algorithmic regimes rather than two minor implementation variants. One can work directly with the raw measurement record and keep the memory footprint low, or one can absorb the combinatorics of all past tuples into dense state variables and thereby obtain a per-shot cost independent of the shot count. In the reconstruction-free regime, a fresh sample participates in $\binom{N-1}{r-1}$ new $r$-tuples, so dependence on the history is unavoidable. In the reconstruction-based regime, that history dependence is compressed into the matrices $A_r^{(N)}$, at the price of $O(md^2)$ memory.
The present work changes the arithmetic inside this second regime. Both the memory scaling and the estimator remain exactly the same.

\subsection{The fixed-memory online recurrence}

The reconstruction-based estimator of \cite{marso2026online} rewrites the same $U$-statistic into a fixed-memory recurrence. For integers $N\ge 0$ and $r\ge 0$, define
$
A_0^{(N)}=I_d
$
and, for $r\ge 1$,
\begin{equation}
A_r^{(N)}
=
\sum_{1\le s_1<\cdots<s_r\le N}
S_{s_1}S_{s_2}\cdots S_{s_r},
\label{eq:Ar}
\end{equation}
with the convention
$
A_r^{(N)}=0
$
when $r>N$.
The recurrence follows by partitioning the tuples in Eq.~\eqref{eq:Ar} into two disjoint classes:
those that do not use the newest shot $N$ and
those whose last index is $N$.
The contribution of the first class is exactly $A_r^{(N-1)}$. For the second class, the first $r-1$ indices range over all increasing $(r-1)$-tuples drawn from $\{1,\ldots,N-1\}$, so their total contribution is
\[
\sum_{1\le s_1<\cdots<s_{r-1}\le N-1}
S_{s_1}\cdots S_{s_{r-1}}S_N
=
A_{r-1}^{(N-1)}S_N.
\]
Adding the two parts yields
\begin{equation}
A_r^{(N)}=A_r^{(N-1)}+A_{r-1}^{(N-1)}S_N,
\qquad 1\le r\le N.
\label{eq:recurrence}
\end{equation}
Taking traces in Eq.~\eqref{eq:Ar} gives
\[
\tr(A_r^{(N)})
=
\sum_{1\le s_1<\cdots<s_r\le N}\tr(S_{s_1}\cdots S_{s_r}),
\]
and comparing with Eq.~\eqref{eq:ustat} yields
$
\widehat p_r^{(N)}=\tr(A_r^{(N)}) / \binom{N}{r}.
$
Thus the whole online state of the estimator is captured by the $m$ matrices $A_1^{(N)},\ldots,A_m^{(N)}$ together with the current shot count $N$. This is precisely why the recurrence is attractive in laboratory settings, as the controller only needs to store the current accumulated matrices rather than the combinatorially large list of all past tuples.

\subsection{Previous arithmetic cost}

The recurrence in Eq.~\eqref{eq:recurrence} is ideal from the memory viewpoint since the full shot history never needs to be retained. The cost bottleneck is the matrix product $A_{r-1}^{(N-1)}S_N$. If one evaluates that product as a generic dense $d\times d$ multiplication, then each order $r$ requires one dense matrix multiplication and one dense matrix addition. In the standard arithmetic model this means
$
O(d^3)+O(d^2)=O(d^3)
$
operations for one order and therefore
$
O(md^3)
$
operations per shot in total. The memory is
$
O(md^2),
$
because one stores $m$ dense $d\times d$ matrices. This is the reconstruction-based baseline that we accelerate.

\section{A faster update via tensorized right multiplication}

\subsection{The structural observation}

At first glance, the matrices $A_r^{(N)}$ defined in Eq.~\eqref{eq:Ar} appear hopelessly dense, and after a few shots they typically are. The online recurrence contains a product
$
A_{r-1}^{(N-1)}S_N
$
with a dense left factor and a newly arrived right factor. The dense left factor carries no useful tensor-factorized form. The right factor does: Eq.~\eqref{eq:factorized} shows that
$
S_N=G_{N,1}\otimes G_{N,2}\otimes\cdots\otimes G_{N,n}.
$

For each qubit $j$, define
$
R_{N,j}=I_{2^{j-1}}\otimes G_{N,j}\otimes I_{2^{n-j}}.
$
Because the $R_{N,j}$ act on different tensor factors, they commute and satisfy
\begin{equation}
R_{N,1}R_{N,2}\cdots R_{N,n}=S_N.
\label{eq:SNproduct}
\end{equation}
Hence any product $MS_N$ may be evaluated as
$
MS_N=((\cdots((MR_{N,1})R_{N,2})\cdots)R_{N,n}).
$
The question is therefore reduced to the cost of multiplying a dense matrix by one local factor $R_{N,j}$.

\subsection{Relation to prior tensor-contraction methods}

The earlier trace-polynomial literature already showed that randomized-measurement witnesses can often be evaluated without explicitly materializing exponentially large tensor products \cite{rico2024entanglement,rico2025state}. Those results, however, do not settle the present online problem. In the offline trace-polynomial setting, one is free to organize the full contraction pattern globally. By contrast, the online recurrence starts shot $N$ from dense matrices $A_r^{(N-1)}$ that already encode all previous data. In general those accumulated matrices have no exact tensor-factorized representation that survives future updates.

The asymmetry of the online product is therefore crucial. The left factor is dense and unstructured, but the fresh right factor still has the exact tensor form of Eq.~\eqref{eq:factorized}. The key observation is that this one-sided structure is already enough. The column-pair sweep below is precisely the mechanism that converts that asymmetric tensor structure into an arithmetic saving for the online update.

\subsection{Cost of one local sweep}

Fix $j\in[n]$ and write
$
G_{N,j}=
\left(
\begin{smallmatrix}
\alpha&\beta\\
\gamma&\delta
\end{smallmatrix}
\right).
$
Label the columns of a $d\times d$ matrix $M$ by $n$-bit strings in lexicographic order. For each $(n-1)$-bit string $u$, let $u^{(0)}$ and $u^{(1)}$ denote the two $n$-bit strings obtained by inserting $0$ or $1$ at position $j$. These label a pair of columns that differ only in bit $j$. Every column belongs to exactly one such pair, so there are exactly $d/2$ disjoint pairs.

Right multiplication by $R_{N,j}$ acts independently on each pair, because $R_{N,j}$ is the identity on all tensor factors other than qubit $j$. This can be seen directly at the level of matrix entries. Let $x,y\in\{0,1\}^n$ label computational-basis strings, and let $x_{\setminus j}$ denote the $(n-1)$-bit string obtained by deleting the $j$th bit of $x$. Then
\[
(R_{N,j})_{x,y}
=
\mathbf{1}\{x_{\setminus j}=y_{\setminus j}\}\,(G_{N,j})_{x_j,y_j}.
\]
In other words, $R_{N,j}$ only couples basis labels that agree everywhere except possibly at qubit $j$, and on each such two-dimensional subspace it acts exactly as the $2\times2$ matrix $G_{N,j}$. Consequently, for every fixed pair $u^{(0)},u^{(1)}$ of columns, the output column indexed by $u^{(a)}$ depends only on the two input columns indexed by the same pair:
\[
(MR_{N,j})_{:,u^{(a)}}
=
\sum_{b\in\{0,1\}} M_{:,u^{(b)}}(G_{N,j})_{b,a},
\qquad a\in\{0,1\}.
\]
Writing this relation out explicitly gives
\begin{equation}
\begin{aligned}
(MR_{N,j})_{:,u^{(0)}}&=\alpha\,M_{:,u^{(0)}}+\gamma\,M_{:,u^{(1)}},\\
(MR_{N,j})_{:,u^{(1)}}&=\beta\,M_{:,u^{(0)}}+\delta\,M_{:,u^{(1)}}.
\end{aligned}
\label{eq:pairsweep}
\end{equation}
Equivalently, if one groups those two columns into a single $d\times 2$ block, then the update is obtained by right-multiplying that block by the local matrix $G_{N,j}$. Thus a single pair update multiplies a $d\times 2$ block by a $2\times2$ matrix. Entrywise, each of the $d$ rows requires only a constant number of scalar operations, so one pair costs $O(d)$ work. Since there are $d/2$ independent pairs, a full sweep over all columns costs
$
O(d^2).
$
No approximation enters here. The sweep is simply the block decomposition of the exact product $MR_{N,j}$. Applying this process successively for $j=1,\ldots,n$ and using Eq.~\eqref{eq:SNproduct} yields an exact multiplication by $S_N$ in
$
O(nd^2)=O(d^2\log d)
$
arithmetic operations.

\begin{algorithm}[h]
\caption{Fast online PT-moment update for all $m$}
\label{alg:fast-online}
\begin{algorithmic}[1]
\REQUIRE Moment order $m$ and subsystem $B\subseteq[n]$
\STATE Initialize $A_0\leftarrow I_d$ and $A_r\leftarrow 0$ for $r=1,\ldots,m$
\FOR{$N=1,2,3,\ldots$}
\STATE Receive the new basis string $P_N=(P_{N,1},\ldots,P_{N,n})$ and outcome string $b_N=(b_{N,1},\ldots,b_{N,n})$
\FOR{$j=1$ to $n$}
\STATE Set $\chi_{N,j}\leftarrow 1$ if $j\in B$ and $P_{N,j}=Y$, otherwise $\chi_{N,j}\leftarrow 0$
\STATE Set $G_{N,j}\leftarrow \frac12 I+\frac32(-1)^{b_{N,j}\oplus\chi_{N,j}}P_{N,j}$
\ENDFOR
\FOR{$r=\min\{N,m\},\min\{N,m\}-1,\ldots,1$}
\STATE $T\leftarrow A_{r-1}$
\FOR{$j=1$ to $n$}
\STATE Replace $T$ by $TR_{N,j}$ using the column-pair sweep in Eq.~\eqref{eq:pairsweep}
\ENDFOR
\STATE $A_r\leftarrow A_r+T$
\ENDFOR
\IF{$N\ge m$}
\STATE Output $\widehat p_m^{(N)}=\tr(A_m)/\binom{N}{m}$
\ENDIF
\ENDFOR
\end{algorithmic}
\end{algorithm}

The resulting procedure is given in Algorithm~\ref{alg:fast-online}. In this algorithm, the descending update order in $r$ is essential, because Eq.~\eqref{eq:recurrence} uses the \emph{old} matrix $A_{r-1}^{(N-1)}$.

\begin{remark}
Fix an $n$-qubit state $\rho$, a bipartition $A|B$, and a moment order $m\ge 1$. After processing $N$ shots, Algorithm~\ref{alg:fast-online} stores the matrices $A_r^{(N)}$ of Eq.~\eqref{eq:Ar} for every $0\le r\le m$. Consequently, whenever $N\ge m$ the output of the algorithm is exactly the unbiased estimator in Eq.~\eqref{eq:ustat}. The memory usage is $O(md^2)$, and the arithmetic cost is
$
O(md^2\log d)=O(mn4^n)
$
per shot, equivalently $O(Nmd^2\log d)$ after $N$ shots.
\end{remark}

\section{A faster update for the second PT moment}
\label{sec:special-m2}

For the second PT moment, and in particular for the purity when $B=\varnothing$, one can do even better than Algorithm~\ref{alg:fast-online}. The key observation is that for $m=2$ the estimator can be recovered from $A_1^{(N)}$ alone, and $A_1^{(N)}$ admits an $O(d)$-time update in the Pauli basis.

Recall from Eq.~\eqref{eq:Ar} that
$
A_1^{(N)}=\sum_{t=1}^N S_t.
$
Therefore
\[
\bigl(A_1^{(N)}\bigr)^2=\sum_{t=1}^N S_t^2+\sum_{\substack{s,t=1\\ s\ne t}}^N S_sS_t.
\]
Taking traces and using cyclicity of the trace gives
\[
\sum_{1\le s<t\le N}\tr(S_sS_t)
=
\frac12\left(\tr\bigl[(A_1^{(N)})^2\bigr]-\sum_{t=1}^N \tr(S_t^2)\right).
\]
Since $\binom{N}{2}=N(N-1)/2$, Eq.~\eqref{eq:ustat} becomes
\begin{equation}
\widehat p_2^{(N)}
=
\frac{\tr\bigl[(A_1^{(N)})^2\bigr]-\sum_{t=1}^N \tr(S_t^2)}{N(N-1)}.
\label{eq:p2-special-identity}
\end{equation}

It remains to compute the two ingredients on the right-hand side. The second term is simple because each local factor
$
G_{t,j}=\frac12 I+\frac32(-1)^{b_{t,j}\oplus\chi_{t,j}}P_{t,j}
$
satisfies
\begin{align*}
G_{t,j}^2
&=
\frac14\Bigl(I+6(-1)^{b_{t,j}\oplus\chi_{t,j}}P_{t,j}+9P_{t,j}^2\Bigr)\\
&=
\frac52 I+\frac32(-1)^{b_{t,j}\oplus\chi_{t,j}}P_{t,j},
\end{align*}
because $P_{t,j}^2=I$. Since each Pauli matrix is traceless,
$
\tr(G_{t,j}^2)=5.
$
Using the factorization $S_t=\bigotimes_{j=1}^n G_{t,j}$ from Eq.~\eqref{eq:factorized} and multiplicativity of the trace on tensor products gives
\begin{equation}
\tr(S_t^2)=5^n
\qquad\text{for every }t.
\label{eq:p2-special-shotnorm}
\end{equation}
Substituting Eq.~\eqref{eq:p2-special-shotnorm} into Eq.~\eqref{eq:p2-special-identity} yields
\begin{equation}
\widehat p_2^{(N)}
=
\frac{\tr\bigl[(A_1^{(N)})^2\bigr]-N5^n}{N(N-1)}.
\label{eq:p2-special-est}
\end{equation}

To update $\tr\bigl[(A_1^{(N)})^2\bigr]$ in $O(d)$ time, we switch only this special case to the Pauli basis. Let $\mathcal{P}_n$ denote the set of all $n$-qubit Pauli strings, so that
$
\tr(QQ')=d\,\delta_{Q,Q'}
$
for all
$
Q,Q'\in\mathcal{P}_n.
$
Write
$
A_1^{(N)}=\sum_{Q\in\mathcal{P}_n}u_Q^{(N)}Q,
$
$
u^{(N)}:=(u_Q^{(N)})_{Q\in\mathcal{P}_n},
$
and
$
\nu^{(N)}=\sum_{Q\in\mathcal{P}_n}\bigl(u_Q^{(N)}\bigr)^2,
$
with
$
u_Q^{(N)}=d^{-1}\tr(QA_1^{(N)}).
$
Because $A_1^{(N)}$ is Hermitian and each $Q\in\mathcal{P}_n$ is Hermitian, all coefficients $u_Q^{(N)}$ are real. Therefore
$
\tr\bigl[(A_1^{(N)})^2\bigr]
=
d\sum_{Q\in\mathcal{P}_n}\bigl(u_Q^{(N)}\bigr)^2
=
d\,\nu^{(N)}.
$

Now consider the newly arrived snapshot $S_N$. Set
$
\xi_{N,j}=(-1)^{b_{N,j}\oplus\chi_{N,j}}.
$
For each subset $U\subseteq[n]$, let $P_{N,U}$ be the Pauli string that equals $P_{N,j}$ on qubits in $U$ and $I$ on qubits outside $U$. Expanding Eq.~\eqref{eq:factorized} gives
\begin{align}
S_N
&=
\sum_{U\subseteq[n]}v_N(U)P_{N,U},\nonumber\\
v_N(U)&=\left(\frac12\right)^{n-|U|}\left(\frac32\right)^{|U|}\prod_{j\in U}\xi_{N,j}.
\label{eq:p2-special-support}
\end{align}
Different subsets $U$ produce different Pauli strings, because each qubit contributes either $I$ or the fixed axis $P_{N,j}$. Hence $S_N$ has exactly $2^n=d$ nonzero Pauli coefficients.

Let $v^{(N)}=(v_Q^{(N)})_{Q\in\mathcal{P}_n}$ denote the sparse coefficient vector of $S_N$ from Eq.~\eqref{eq:p2-special-support}. Since
$
A_1^{(N)}=A_1^{(N-1)}+S_N,
$
the coefficient vectors satisfy
$
u^{(N)}=u^{(N-1)}+v^{(N)}.
$
Therefore
$
\nu^{(N)}
=
\nu^{(N-1)}
+
2\sum_{Q\in\mathcal{P}_n}u_Q^{(N-1)}v_Q^{(N)}
+
\sum_{Q\in\mathcal{P}_n}\bigl(v_Q^{(N)}\bigr)^2.
$
Both sums may be evaluated by iterating only over the $d$-element support of $v^{(N)}$. Applying the same Pauli-orthogonality identity to $S_N$ and using Eq.~\eqref{eq:p2-special-shotnorm} gives
$
\sum_{Q\in\mathcal{P}_n}\bigl(v_Q^{(N)}\bigr)^2
=
\frac{\tr(S_N^2)}{d}
=
\frac{5^n}{d}.
$
Thus one shot can be processed by a single pass over the nonzero Pauli coefficients of $S_N$.

\begin{algorithm}[h]
\caption{Fast online PT-moment update for $m=2$}
\label{alg:fast-online-m2}
\begin{algorithmic}[1]
\REQUIRE Subsystem $B\subseteq[n]$
\STATE Initialize $u_Q\leftarrow 0$ for every $Q\in\mathcal{P}_n$ and $\nu\leftarrow 0$
\FOR{$N=1,2,3,\ldots$}
\STATE Receive the new basis string $P_N=(P_{N,1},\ldots,P_{N,n})$ and outcome string $b_N=(b_{N,1},\ldots,b_{N,n})$
\FOR{$j=1$ to $n$}
\STATE Set $\chi_{N,j}\leftarrow 1$ if $j\in B$ and $P_{N,j}=Y$, otherwise $\chi_{N,j}\leftarrow 0$
\STATE Set $\xi_{N,j}\leftarrow (-1)^{b_{N,j}\oplus\chi_{N,j}}$
\ENDFOR
\STATE $\Delta\leftarrow 0$
\FOR{each $(Q,v)\in\operatorname{supp}_{\mathrm P}(S_N)$}
\STATE $\Delta\leftarrow \Delta+u_Qv$
\STATE $u_Q\leftarrow u_Q+v$
\ENDFOR
\STATE $\nu\leftarrow \nu+2\Delta+5^n/d$
\IF{$N\ge 2$}
\STATE Output $\widehat p_2^{(N)}=(d\nu-N5^n)/(N(N-1))$
\ENDIF
\ENDFOR
\end{algorithmic}
\end{algorithm}

For implementation, index the dense array $(u_Q)_{Q\in\mathcal{P}_n}$ by the base-$4$ encoding of the Pauli strings with $I\mapsto 0$, $X\mapsto 1$, $Y\mapsto 2$, and $Z\mapsto 3$. Let $\operatorname{supp}_{\mathrm P}(S_N)$ denote the ordered list of the $d$ pairs $(P_{N,U},v_N(U))$ from Eq.~\eqref{eq:p2-special-support}, arranged in Gray-code order over subsets $U\subseteq[n]$.
Define $\mathrm{enc}(X)=1$, $\mathrm{enc}(Y)=2$, and $\mathrm{enc}(Z)=3$. If qubit $j$ is toggled into the subset, the current coefficient is multiplied by $3\xi_{N,j}$ and the array index is increased by $\mathrm{enc}(P_{N,j})4^{j-1}$. If qubit $j$ is toggled out, the coefficient is multiplied by $\xi_{N,j}/3$ and the same index increment is subtracted.
Hence each emitted pair costs $O(1)$ work. The resulting procedure is given in Algorithm~\ref{alg:fast-online-m2}.

\begin{remark}
The dense coefficient array $(u_Q)_{Q\in\mathcal{P}_n}$ has $4^n=d^2$ real entries, so the memory usage is $O(d^2)$. Each shot touches only the $d=2^n$ nonzero coefficients of $S_N$, and the Gray-code stream emits each of them in $O(1)$ time, so the arithmetic cost is $O(d)$ per shot and $O(Nd)$ after $N$ shots. This improves the $m=2$ specialization of Algorithm~\ref{alg:fast-online} from $O(d^2\log d)$ per shot to $O(d)$ per shot while keeping the same $O(d^2)$ memory.
\end{remark}

\begin{table}[h]
\centering
\caption{Per-shot time and memory comparison for fixed-memory online PT-moment estimation methods.}
\label{tab:comparison}
\begin{tabular}{lcc}
\toprule
Method & Per-shot time & Memory \\
\midrule
Marso \emph{et al.} \cite{marso2026online} & $O(md^3)$ & $O(md^2)$ \\
Alg.~\ref{alg:fast-online} & $O(md^2\log d)$ & $O(md^2)$ \\
Alg.~\ref{alg:fast-online-m2} & $O(d)$ & $O(d^2)$ \\
\bottomrule
\end{tabular}
\end{table}

The comparison is summarized in Table~\ref{tab:comparison}.

\section{Experiments}
We benchmarked the dense implementation of Marso \emph{et al.} \cite{marso2026online} against Algorithms~\ref{alg:fast-online} and \ref{alg:fast-online-m2} for the end-to-end PT-moment estimation pipeline implemented in NumPy. The comparison was performed for $n=9,10,11,12$ qubits and moment orders $m=2,3,4,5$ with $N=10$ shots, where we fixed $B=\{\lfloor n/2\rfloor+1,\ldots,n\}$. Using \(N=10\) provides a representative comparison of the average per-shot runtime over ten samples. To ensure a fair comparison, all experiments were restricted to single-threaded execution (though Algorithm~\ref{alg:fast-online} is well suited to multicore and GPU environments because the \(d/2\) column-pair updates within each local sweep are mutually independent). The results are reported in Table~\ref{tab:runtime_comparison}. As expected, the advantage of Algorithms~\ref{alg:fast-online} and \ref{alg:fast-online-m2} increases with the number of qubits, yielding greater runtime savings at larger system sizes.
At \(n=12\), the observed memory consumption was 1611 MB versus 134 MB for \(m=2\), and 3221 MB versus 2147 MB for \(m=5\), for Marso \emph{et al.} \cite{marso2026online} and our method, respectively.

\begin{table}[h]
\centering
\caption{Runtime comparison (in seconds) for the end-to-end PT-moment estimation pipeline. Bold values indicate the fastest time for each parameter pair $(n, m)$.}
\label{tab:runtime_comparison}
\begin{tabular}{cccccc}
\toprule
$n$ & $m$ & Marso \emph{et al.} \cite{marso2026online} & Alg.~\ref{alg:fast-online} & Alg.~\ref{alg:fast-online-m2} \\
\midrule
\multirow{4}{*}{9}
  & 2 & .1858 & .6207 & $\mathbf{6.355\times10^{-4}}$ \\
  & 3 & \textbf{.3370} & .6272 &  \\
  & 4 & \textbf{.4919} & .6880 &  \\
  & 5 & \textbf{.6468} & .7339 &  \\
\midrule
\multirow{4}{*}{10}
  & 2 & 1.291 & 1.074 & $\mathbf{8.474\times10^{-4}}$ \\
  & 3 & 2.471 & \textbf{1.306} &  \\
  & 4 & 3.633 & \textbf{1.522} &  \\
  & 5 & 4.852 & \textbf{1.695} &  \\
\midrule
\multirow{4}{*}{11}
  & 2 & 9.836 & 2.551 & $\mathbf{2.441\times10^{-3}}$ \\
  & 3 & 19.44 & \textbf{3.440} &  \\
  & 4 & 28.84 & \textbf{4.191} &  \\
  & 5 & 37.86 & \textbf{4.809} &  \\
\midrule
\multirow{4}{*}{12}
  & 2 & 76.18 & 9.591 & $\mathbf{6.564\times10^{-3}}$ \\
  & 3 & 151.6 & \textbf{13.46} &  \\
  & 4 & 224.9 & \textbf{16.83} &  \\
  & 5 & 302.0 & \textbf{19.68} &  \\
\bottomrule
\end{tabular}
\end{table}

\section{Limitations}
The algorithm still stores $m$ dense $d\times d$ matrices, so it is naturally a moderate-system method. If $d^2$ itself does not fit in memory, no dense-matrix recurrence of this type will be satisfactory. Our contribution is therefore not to remove the exponential barrier altogether, but to eliminate the dominant time bottleneck in the regime where the reconstruction-based strategy is otherwise attractive.
Secondly, we do not address the complementary reconstruction-free estimator of \cite{marso2026online}, whose advantage is low memory rather than high data locality. Whether a similarly clean acceleration exists in that regime remains open. So do lower bounds for fixed-memory online PT-moment estimation, which would clarify how close the $O(md^2\log d)$ update is to optimal.
Lastly, the argument is tailored to local Pauli classical shadows and to PT-moment estimation under partial transpose. It does not automatically imply the same improvement for arbitrary measurement ensembles or for arbitrary nonlinear shadow functionals. Extending the one-sided factor exploitation principle beyond the Pauli/PT setting is an interesting direction for future work.

\section{Conclusion}

We have presented a faster implementation of the reconstruction-based online PT-moment estimator for local Pauli classical shadows. The improvement comes from evaluating the product with the partially transposed snapshot through $n$ local column-pair sweeps instead of dense matrix multiplication.
For the case $m=2$, a Pauli basis update reduces the per-shot cost further to $O(d)$.
This yields an improved arithmetic complexity while preserving the memory usage. Because PT moments are a central ingredient in entanglement certification, the result provides a practically relevant acceleration of the online classical post-processing loop in shadow-based experiments.

\section*{Acknowledgment}
This work is in part supported by the National Research Foundation of Korea (NRF, RS-2024-00451435 (20\%), RS-2024-00413957 (20\%)), Institute of Information \& communications Technology Planning \& Evaluation (IITP, RS-2025-02305453 (15\%), RS-2025-02273157 (15\%), RS-2025-25442149 (15\%), RS-2021-II211343 (15\%)) grant funded by the Ministry of Science and ICT (MSIT), Institute of New Media and Communications (INMAC), and the BK21 FOUR program of the Education, Artificial Intelligence Graduate School Program (Seoul National University), and Research Program for Future ICT Pioneers, Seoul National University in 2026.


\bibliographystyle{IEEEtran}
\bibliography{references}

\end{document}